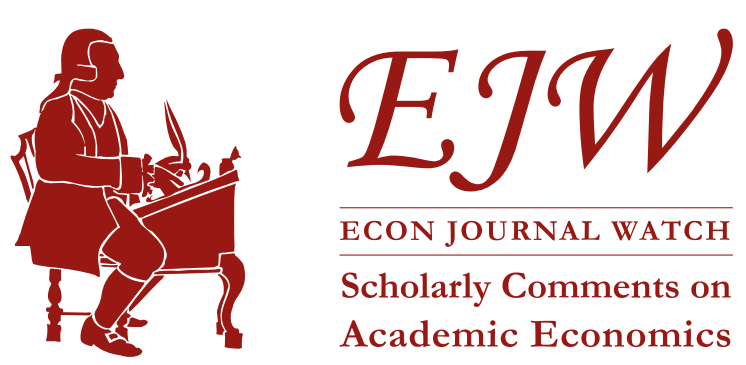

# Power Analysis is Essential: High-Powered Tests Suggest Minimal to No Effect of Rounded Shapes on Click-Through Rates[1]

Ron Kohavi[2], Jakub Linowski[3], Lukas Vermeer[4], Fabrice Boisseranc[5], Joachim Furuseth[6], Andrew Gelman[7], Guido Imbens[8], Ravikiran Rajagopal[9]

Online controlled experiments, or A/B tests, are commonly used to evaluate ideas and their implementations in websites and applications (G. Z. Georgiev 2019; Kohavi, Tang, and Xu 2020; Luca and Bazerman 2020; Thomke 2020; Weigel 2024). Large companies run tens of thousands of A/B tests per year. The thirteen organizations that attended the Practical Online Controlled Experiments Summit in 2018 "tested more than one hundred thousand experiment treatments last year"

[1] *Acknowledgements*: We thank Brian Nosek and Daniel Lakens for their encouragement to replicate the initial study and submit a paper. We thank David Allison, Andrew Brown, Deborah O'Malley, Uri Simonsohn, and Colby Vorland for their feedback and suggestions.

[2] Sole Proprietor.
[3] Founder, Linowski Interaction Design, Inc.
[4] Independent researcher.
[5] Customer Success Lead, Kameleoon.
[6] CRO manager, Coop Norway.
[7] Professor, Department of Statistics and Professor, Department of Political Science, Columbia University, New York.
[8] Graduate School of Business and Department of Economics, Stanford University, SIEPR, NBER, Stanford, CA.
[9] Director of Product Management, United Parks & Resorts.

(Gupta et al. 2019). Six years later, Microsoft alone now runs 100,000 A/B tests annually (Kumar Bajpai et al. 2025).

Several of the present authors have been intimately involved in large-scale A/B tests at multiple large companies and have reviewed thousands of experiments, which included changes to the user interface, element colors, and shapes. We have a good sense of the lifts (relative treatment effects), which are typically under one percent, and rarely over two or three percent. It was therefore with surprise and great skepticism that we viewed the results of the *Journal of Consumer Research* article, "Curvy Digital Marketing Designs: Virtual Elements with Rounded Shapes Enhance Online Click-Through Rates" by Dipayan Biswas, Annika Abell, and Roger Chacko (2023). Their Study 2—the only valid A/B test among all the studies described in that paper—reported a 55% increase in click-through rate by changing web elements from square to curved.

Biswas, Abell, and Chacko (2023)—henceforth abbreviated BAC—also reported a second "Field Experiment" in Study 1, but the authors clearly stated that "assignment of consumers to ads is not randomized" by Google Ads. Indeed, Google Ads intentionally target users who are more likely to click on a given ad, to make effective use of the ad budget. Such divergent delivery should not be used to deduce any causal effects as if it were a randomized controlled experiment (Braun and Schwartz 2025). In "On the persistent mischaracterization of Google and Facebook A/B tests: How to conduct and report online platform studies" (Boegershausen et al. 2025, 887), the authors warn about Google Ads and write that "we purposefully use the term online platform 'studies' to avoid giving the false impression that they are 'experiments' or true A/B tests with proper randomization to different treatments…online platform studies employ post-randomization targeting algorithms that prevent clean random assignment of participants…Lacking internal validity, online platform studies should not be considered as randomized experiments."

Furthermore, Study 1 suffers from a large Sample Ratio Mismatch (SRM) (Kohavi and Thomke 2017; Fabijan et al. 2019), which is a strong deviation of the ratio between impressions in the two variants from the experimental design. While the design was made for an even split of 50:50, the curved ads received 65,495 impressions and the sharp-angled ads received 83,331 impressions, a ratio of 44:56. The probability of such a ratio, or a more extreme ratio, is about $10^{-460}$, with a z-score of 46. For comparison, the probability of randomly picking a specific atom out of all the atoms in the observable universe is about $10^{-80}$. When such an unlikely event occurs, the experiment is deemed not trustworthy, which is why it is a common guardrail in experimentation platforms (Kohavi, Tang, and Xu 2020).

Given the lack of randomization and the extreme sample ratio mismatch observed in Study 1, we were surprised that this analysis was presented as a field experiment contributing causal evidence.

BAC includes two additional small lab studies (Study 3 and Study D) that evaluate the impact of rounded shapes on click-through rate and can thus be compared. Lab studies typically have worse external generalizability (Lynch 1982), but we recognize that the BAC authors designed them well to hide the underlying metrics being evaluated and improve their external generalizability. If their lifts are expected to generalize to real-world settings, they are severely under-powered, as shown below.

We conducted three experiments comparing square and rounded buttons: one at SeaWorld® Orlando and the others on two different websites that are part of Coop, Norway's 15th largest company.

The experiments were run as part of the community project *Trustworthy A/B Patterns* (Kohavi, Linowski, and Vermeer 2024). We did not pre-register the experiment publicly (e.g., on OSF or AsPredicted.org), as such pre-registration would require disclosure of sensitive client information (e.g., traffic) that was not approved by our NDA. Note that BAC's Study 2 was not publicly pre-registered either, likely due to similar NDA restrictions. As a replication experiment, most elements of the experiment plan and analysis were pre-defined: the intervention (square vs. rounded corners in visual elements), the primary metric (click-through rate), no filtering of data, and no transformations (matching the original experiment). We also publicly called out the MDE (Minimum Detectable Effect) as relative 2% lift before the Coop experiments started. When the results from the Coop experiments were not statistically significant, we then updated the target MDE to 0.5% before the SeaWorld experiment started (Kohavi, Linowski, and Vermeer 2024). For all experiments, we determined the duration pre-experiment with the participating sites based on the desired MDE, thus restricting any undisclosed flexibility that could increase the false-positive rate (Simmons, Nelson, and Simonsohn 2011). Unlike our experiments, BAC did not discuss an MDE or include any power calculations.

The American Economic Association's Registry for Randomized Controlled Trials has reviewed our three experiments, including time-stamped versions of our public documents posted on the web, and has approved them as registered trials (Kohavi, Linowski, and Vermeer 2025; Kohavi, Linowski, Vermeer, and Furuseth 2025; Kohavi, Linowski, Vermeer, and Rajagopal 2025).

BAC's Study 2 was based on 919 visits and reported an effect size of 55%. Simonsohn (2015) suggests that replication sample size be 2.5 times the original sample. Our replications had over two thousand times more users than in the initial study. The three experiments we conducted had over 2.8 million users, 2.2 million users, and 1.9 million users with estimated effect sizes of 0.16% (p-value 0.20), 0.29% (p-value 0.60), and 0.73% (p-value 0.09).

We do not think focusing on the p-values is especially useful here. Even if this change did represent a true effect, these results are two orders of magnitude smaller than reported by BAC. Demidenko (2016, 33) shares an example of an anti-obesity drug, which reduces weight statistically significantly but the effect size of which is one pound, or 0.4% of a 250-pound person; he concludes, "I doubt that anybody would buy this drug." However, if the effect size was 100 times larger, you would have a blockbuster weight-loss drug, better than recent industry break-through drugs, such as Semaglutide marketed as Wegovy® (Wilding et al. 2021) and Tirzepatide marketed as Zepbound® (Jastreboff et al. 2022). The same is true here: What was claimed to be a striking finding with potentially wide-ranging implications is likely a tiny effect in reality. It is not clear that organizations should prioritize this idea relative to many other ideas that have a similar or greater expected gain. With such a small estimated effect size, most organizations would not even have enough statistical power to evaluate whether the change increases click-through rate, let alone mention purchases or revenue, which our large sites lacked sufficient power to evaluate.

We end this section with an example that comes from the book *Active Statistics* (Gelman and Vehtari 2024), where an experiment changed the call for donation box from square to round corners, closely matching the treatment discussed here. It was run by the Wikimedia Foundation, the organization that funds Wikipedia. The estimated treatment effect was –15%, which the people at Wikimedia felt was implausibly large. They contacted an outside statistician, who noticed from the output of the A/B testing analysis software that the assignments were unbalanced. It turned out that, due to a bug in the treatment assignment or data recording process, the fraction of data in the treatment group was not the desired 50% throughout the testing period, but rather fluctuated between 50% and 100%, creating a sample ratio mismatch (Kohavi and Thomke 2017; Fabijan et al. 2019), the same problem that invalidated BAC's Study 1. The key here is that an extreme result like –15% is so uncommon in the industry that it was considered implausible. 55% would therefore be even more implausible.

## Choosing a sample size for an A/B test

Before running an A/B test, one should choose a sample size such that the experiment is sufficiently powered for the research and decision goals of the experiment. Many have called out the importance of sufficiently powering controlled experiments (Brysbaert 2019; Vankov, Bowers, and Munafò 2014; Perugini, Gallucci, and Costantini 2018; Cohen 1992). One of the simplest formulations for the size of each variant when using the industry default of 0.05 for alpha (the p-value threshold for statistical significance) and 80% power (Lehr 1992; van Belle 2008)

is:

$$n \approx \frac{16\sigma^2}{\delta^2} \quad (1)$$

where $\sigma$ is the standard deviation of the measurement within each group and $\delta$ is the minimum detectable effect (MDE). While $\sigma$ is usually easy to estimate based on company data, the MDE is typically set based on a lower bound determined by historical estimates of similar experiments. This is where online controlled experiments, or A/B tests, diverge from classical psychological studies. In psychology, Cohen's *d* of 0.2 is considered small (Brysbaert 2019), but in online settings, a 5% improvement to 5% conversion (0.25% absolute delta) is considered a large win (see below), and Cohen's *d* and Cohen's *h* for that are about 0.01—tiny. How should one choose a realistic MDE for the power formula? One has to establish some prior distribution on the expected effect size and then pick an MDE based on that prior, taking into account the desired minimum effect and trading it off against the experiment duration. Below we share several possible approaches for establishing that prior distribution in online settings: repositories, average effects at large companies, and expert opinions.

## Online Repositories

There are multiple repositories documenting A/B tests and their treatment effect sizes. Below we share the examples closest to the rounded vs. square corners treatment that we could find.

1. GoodUI.org (Linowski 2025) documents A/B testing patterns and currently has over 600 A/B tests categorized into 141 patterns. It is a great source for finding similar patterns to help estimate an expected MDE. There are 77 tests related to buttons on the website.[10] Here are the best three A/B tests we found that impact buttons similarly to BAC:
    a. Test #526 by Metro-CC attempted to make the search bar a higher contrast. It had 896,009 visits and the estimated treatment effect was a 5.6% increase in sales with p-value 0.014.
    b. Test #582 by Metro-CC made the quantity icons more visible. It had 1,242,470 visits and the estimated treatment effect was a 1.8% decrease in sales with p-value 0.27.
    c. Test #310 by Backstage.com changed the "View All Details" button style from a high contrast blue background to a border style. It had 594,997 visits and the estimated treatment effect was a 1.6% increase in sales with p-value 0.68.

[10] All tests can be found here: https://goodui.org/patterns/tag/button/

2. Evidoo (Jongejan 2025) claims to be the world's largest online A/B test database for e-commerce, based on over 20 years of conducted A/B tests. It contains a rich set of 3,000 patterns. Here are the best examples we found that are related to buttons similarly to BAC:

    a. Best practice #97 is based on 23 A/B tests where a Unique Selling Proposition (USP) is added near the add-to-cart-button. It won (statistically significant treatment effect) 39% of the time with an average impact of 6.8%.
    b. Best practice #267 is based on two A/B tests and adds/removes social share buttons. It won 50% of the time with an average impact of 1.2%.
    c. Best practice #268 is based on five A/B tests and adds/removes an Apple Pay button. It won 20% of the time with an average impact of 6.1%.

In all these highly successful patterns, the effects are under 7%, so we believe an appropriate MDE that sets a lower bound for similar patterns should be around 2–3%.

## Average Effects at Large Companies Running A/B Tests

Another approach to choosing a realistic MDE is to look at the average or median of companies running many A/B tests. Here are three examples:

1. Azevedo et al. (2020, Figure 3(a)) report that in thousands of experiments at Microsoft/Bing, the treatment effect is rarely over 0.3%.

2. At Airbnb search relevance, 250 experiments over 1.5 years improved conversion by a cumulative total of 6%. The average successful experiment improved conversion by 0.3% (Kohavi 2023).

3. Analytics Toolkit, a third-party web analytics platform, reported a relative median lift of 0.1% and an average relative lift of 2.1% after reviewing 1,001 experiments (Georgiev 2022).

With such small effect sizes, there is good evidence to choose a small MDE, yet as we show below, BAC's sample size does not even support an MDE of 10%, much higher than the average above.

## Elicitation of Practitioner Expectations

Elicitation of expert judgment can help design choices, such as an MDE (Morgan and Henrion 1990). In December 2024, before any results from our two ongoing Coop experiments were available, we conducted a LinkedIn poll of practitioners with A/B testing experience. Participants were shown a link to BAC's article and could optionally visit the websites to view the variants.[11]

Of 115 respondents, we removed 28 respondents who self-reported to have been involved with ten or fewer A/B tests, leaving us with respondents who had practical experience in 11 or more A/B tests. Figure 1 summarizes the 87 responses of these experienced respondents. The predicted average lift was 0.87%, the median prediction was 0.1%, 17 respondents predicted no effect, and 44 respondents (about 51%) predicted an effect within [–0.5%, 0.5%]. The largest prediction was a treatment effect of 8.2%, still six times smaller than the 55% lift reported by BAC. Because respondents are self-selected and uncalibrated, we use this elicitation only to motivate plausible effect sizes for choosing an MDE, not as evidentiary support about the true causal effect.

**Figure 1:** Expert predictions of the treatment effect for rounded corners

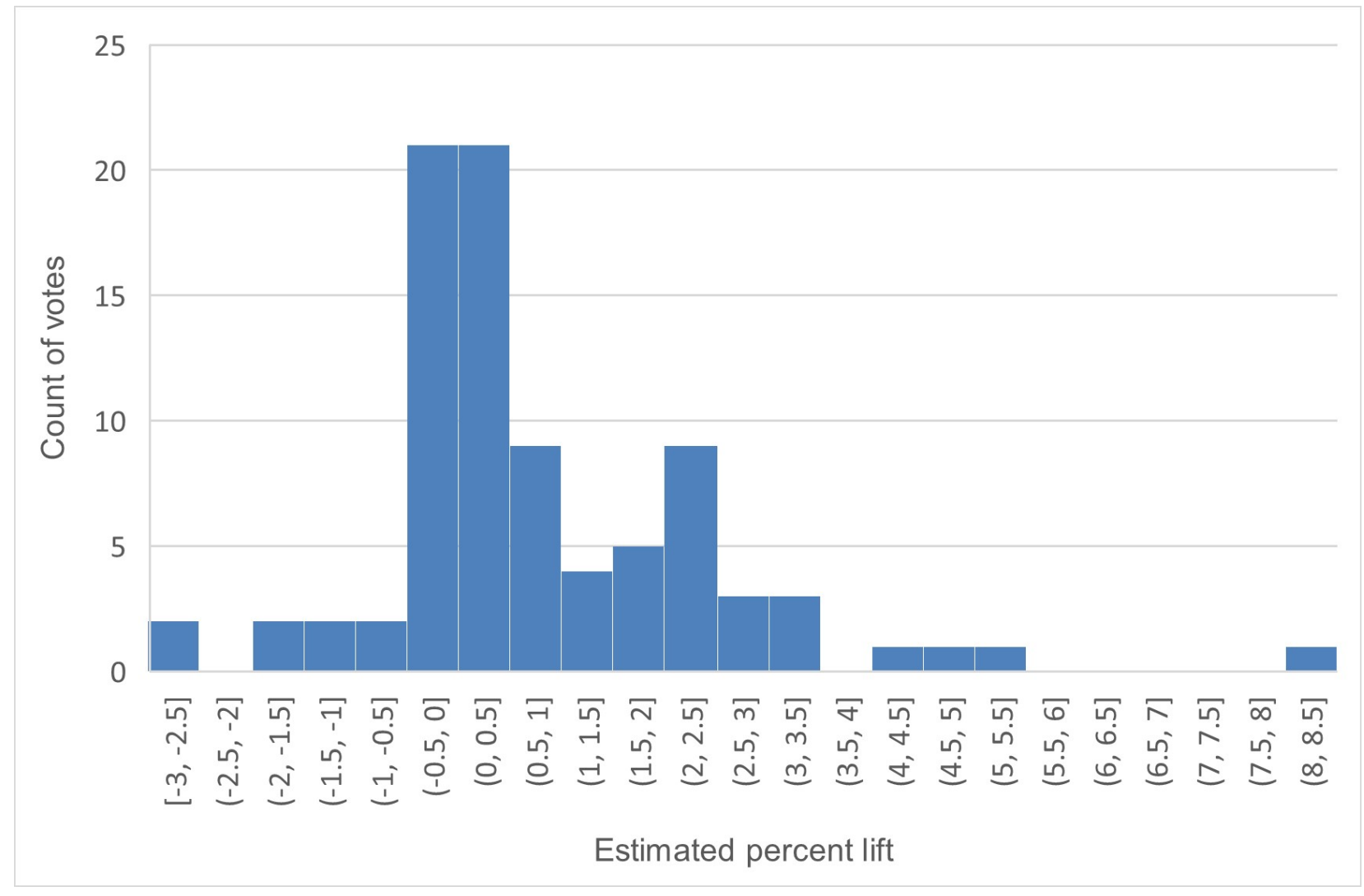

[11] The full poll materials and the question wording are available at the OSF repository https://osf.io/qen6a/files/dvjg7.

### Computing the required sample size to achieve power

In the community project *Trustworthy A/B Patterns* (Kohavi, Linowski, and Vermeer 2024) that called for replicating the square to round corners pattern, an MDE of relative 2% lift was recommended.

The original A/B test reported in BAC had a click-through rate of 7.19% for the control, and hence $\sigma = \sqrt{p(1-p)} = 0.26$. Using an MDE of 2% would require over

$$\frac{16 * 0.06673}{(0.0719 * 2\%)^2} = 516{,}325 \tag{2}$$

users per variant. Even an MDE of 5% would require over 80,000 users per variant, and an MDE of 10% would still require over 20,000 users per variant. The A/B test in BAC had just 474 visits in one variant and 445 visits in the other, demonstrating that it was severely underpowered even if we use a relatively large MDE.

We also note that randomizing an A/B test by users and analyzing by visits can introduce its own biases, as the visits are not independent, requiring an analysis using the Delta method (Deng, Knoblich, and Lu 2018). BAC note that "Google Optimize [the software used to run the A/B test] ensures that individual users would be exposed to only one of the experimental conditions" (Biswas, Abell, and Chacko 2023, 557), which implies that the user, not the visit, was the randomization unit. Our experiments were all randomized *and* analyzed at the user level.

## Details of our high-power experiments

We conducted three experiments comparing square and rounded buttons on the following websites:

1. SeaWorld® Orlando (https://seaworld.com/orlando/), part of United Parks & Resorts Inc., is a global theme park and entertainment company that operates 13 parks in seven markets. SeaWorld® Orlando is a marine-life and thrill-ride theme park located in Orlando, Florida. It hosts over 4 million visitors annually, making it one of the largest theme parks in North America. The Company is a global leader in animal care, managing a major zoological collection and rescuing over 40,000 animals through its SeaWorld® rescue team. The SeaWorld Orlando website allows guests to explore experiences, rides, attractions, and events, as well as plan visits and purchase tickets and add-ons.

2. Obs (https://www.obs.no), Norway's largest hypermarket, with 31 physical stores throughout the country, sells groceries and non-food; the web shop sells non-food items. The product range varies from PlayStation consoles to gardening tools, sporting equipment, and LEGO®.

3. Obs-BYGG (https://www.obsbygg.no), one of Norway's largest DIY (do-it-yourself) chains with 59 stores throughout the country, sells everything you need for your home improvement projects, your garden, or your garage. The product range varies from a broad selection of tools, paint, flooring, windows, doors, and equipment and furniture for your garden.

We now describe the experimental design for the three websites. For SeaWorld, which had square buttons throughout, we were able to programmatically change the buttons to round in the Treatment. This included buttons everywhere, including add-to-cart shown in Figure 2, the popup asking for email address shown in Figure 3, and dates buttons shown in Figure 4. Given the large number of buttons, the click-through rate (CTR) per user (on any button) was over 40%.

The public *Trustworthy A/B Patterns* document initially recommended a 2% MDE back in 2024, but after the results of the two Coop experiments were published in January 2025, the recommendation was updated to 0.5% on May 6, 2025, before the SeaWorld experiment started. Power calculations indicated that the experiment should run for six weeks, which was agreed and documented before the experiment started on May 19, 2025. The experiment ended as planned after 42 days on June 30, 2025.

For Coop, only the add-to-cart button was changed. The historical user to add-to-cart conversion rate was about 10% for Obs and 5.5% for Obs-BYGG. This was due to a different user base and product mix, providing useful natural variability to the evaluation with a relative 80% difference in those conversion rates.

We changed the add-to-cart buttons from round to square. To align with BAC's presentation, we present the results as if the control was square. The square and round variants are shown in Figure 5 and shown zoomed out in Figure 6.

At the time, the public *Trustworthy A/B Patterns* document recommended an MDE of 2%. Power calculations indicated that one experiment should run for slightly under two months and the other for slightly over two months. We agreed to run them both for 16 weeks so that we would have additional power. The Obs experiment started on September 3, 2024, and the Obs-BYGG experiment started September 11, 2024. Both ran for exactly 16 weeks (112 days). At the end of the 16 weeks, there were over 970,000 users in each of the two variants in the Obs site, and over 1,100,000 users in each of the two variants in the Obs-BYGG site.

The Obs and Obs-BYGG experiments had a second treatment implemented, which had slightly rounded corners. For completeness, we share the results in Appendix B. The results are similar in effect size, and neither of the two treatments was statistically significant at the 0.05 level. While they could be viewed as pro-viding additional evidence for a small treatment effect, they are not independent and therefore do not enhance external generalizability, nor do they replicate faithfully the original rounding as implemented in BAC. We believe our approach here to focus on the three independent replications and not claim five replications is conservative.

**Figure 2:** SeaWorld add-to-cart buttons: Control (left) was square, Treatment (right) was round.

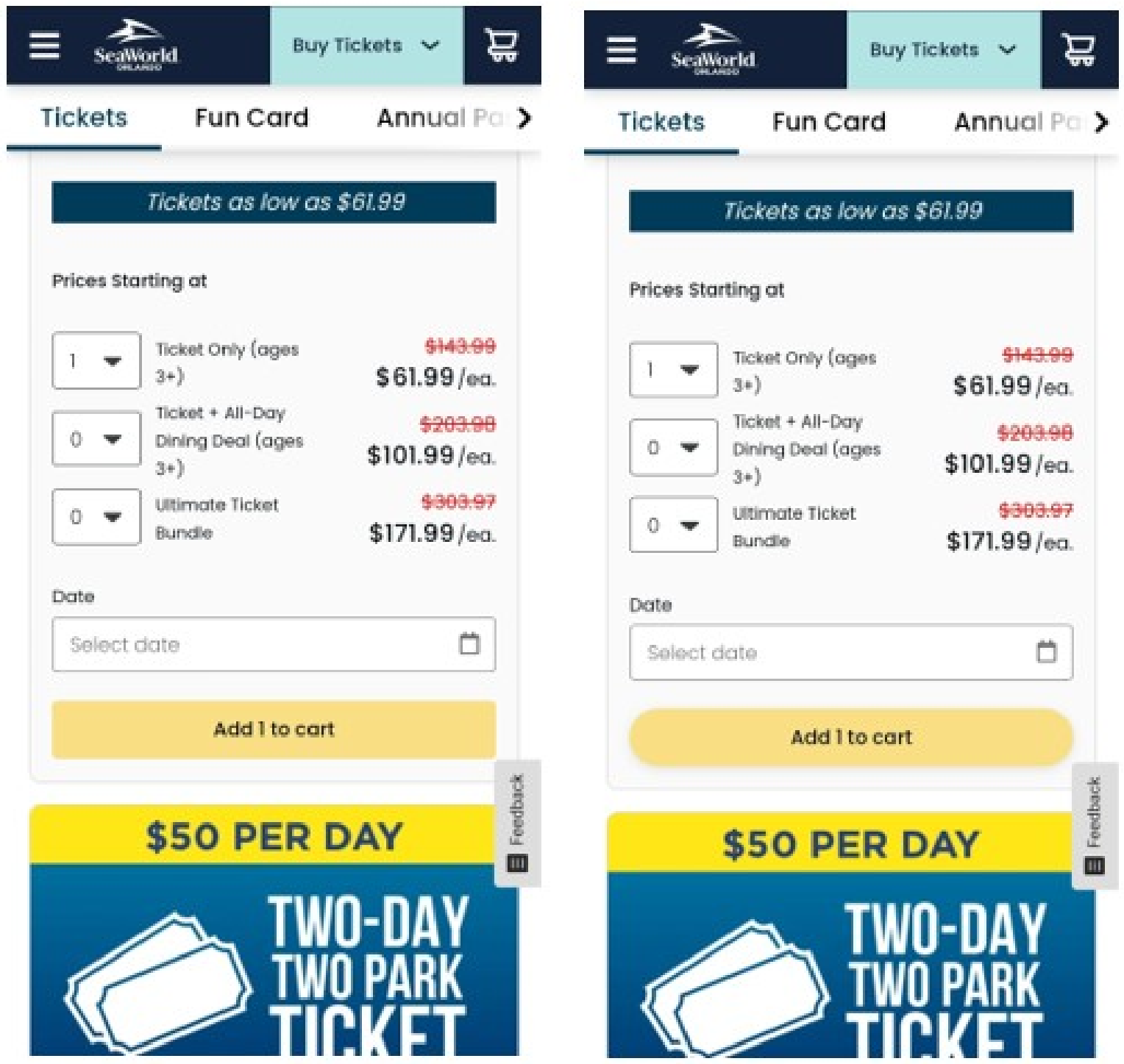

**Figure 3:** SeaWorld popup for new users: Control (left) was square, Treatment (right) was round

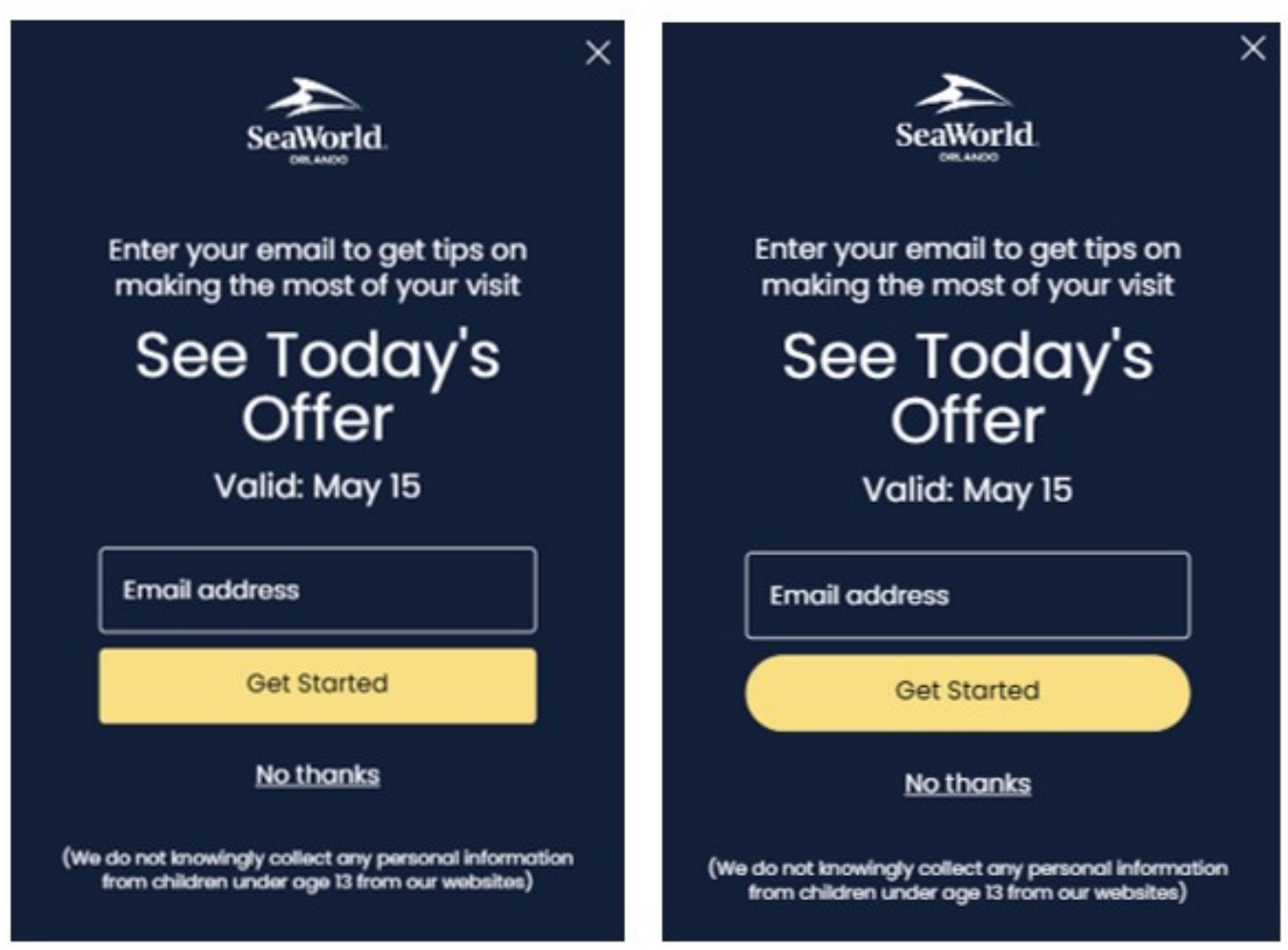

**Figure 4:** SeaWorld select date button for the Dolphin Encounter: Control (left) was square, Treatment (right) was round

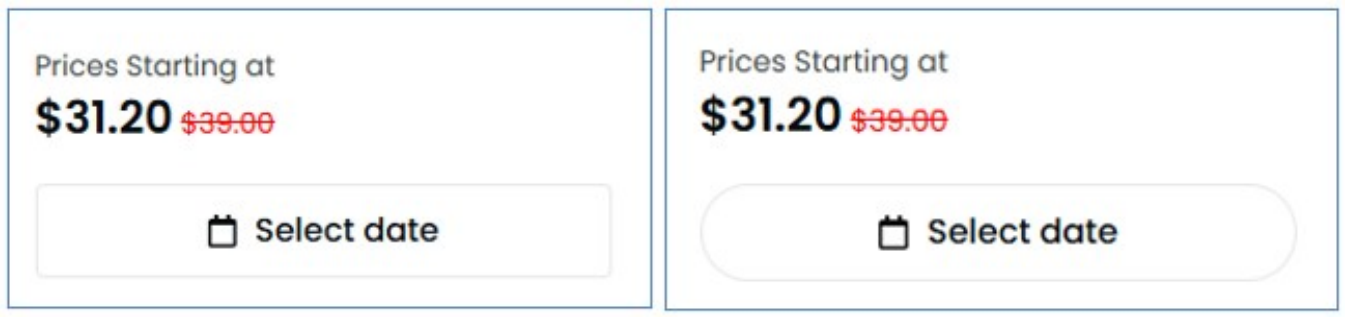

**Figure 5:** Square buttons variant vs. rounded buttons variant

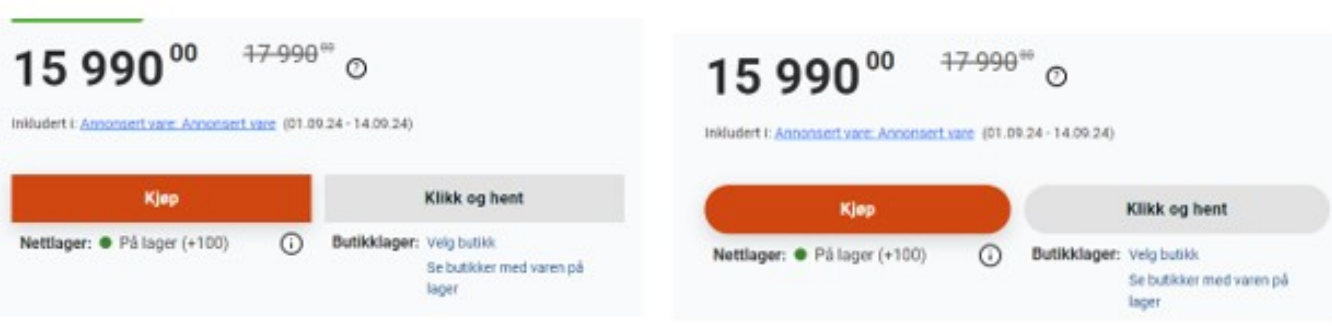

**Figure 6:** Zoomed out variants

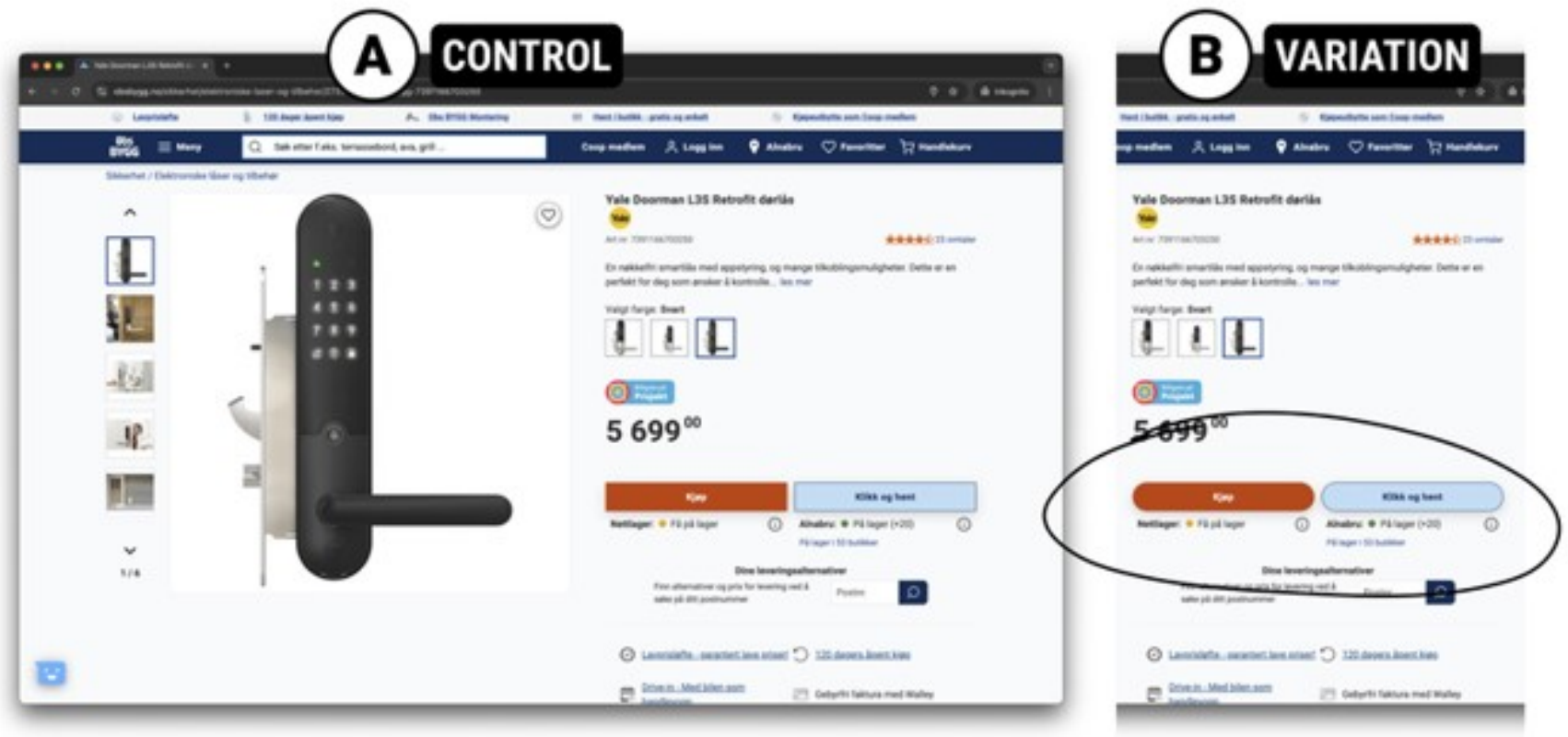

Table 1 below summarizes the results of BAC's Study 2, the two additional lab studies that evaluated click-through rate (Study 3 and Study D), and our three independent replications.

**TABLE 1: Summary of Results**

| Study | N control | N treatment | CTR Control | CTR Treatment | Lift (T-C)/C | p-value |
|---|---|---|---|---|---|---|
| BAC (Study 2) | 445 | 474 | 7.19% | 11.18% | 55.49% | 0.037 |
| BAC (Study 3) | 50 | 54 | 52.00% | 75.93% | 46.01% | 0.011 |
| BAC (Study D) | 32 | 33 | 21.88% | 45.45% | 107.79% | 0.045 |
| SeaWorld® Orlando | 1,448,041 | 1,448,066 | 47.13% | 47.21% | 0.16% | 0.20 |
| Obs-BYGG | 1,126,132 | 1,124,100 | 5.43% | 5.45% | 0.29% | 0.60 |
| Obs | 977,499 | 976,653 | 10.07% | 10.14% | 0.73% | 0.09 |

An Excel spreadsheet with the detailed analysis and full precision computations is available at https://bit.ly/roundedCornersComputations under the meta-analysis tab. The meta-analysis was done in two ways: directly in Excel and using the CMA meta-analysis software (Biostat 2022) and the results match. The meta-analysis combining all three replication experiments results in a (weighted-average) lift of 0.21% with p-value 0.08 (Borenstein et al. 2021). With over 7.1M users in total, we were unable to reject the null hypothesis of no effect.

We have done several robustness checks:

1. The data for SeaWorld was collected by Optimizely, an A/B testing platform.[12] Optimizely analyzes the data using an always-valid-p-value approach (Johari et al. 2017; Johari et al. 2022) that allows peeking at p-values. We pre-declared that we would analyze the experiment using the classical fixed horizon, and our table above reports the p-value of 0.20 based on that. Optimizely's "Statistical Confidence" was 92% (an always-valid-p-value of 0.08). During the experiment, the Confidence varied materially and did not maintain monotonicity, contrary to the documentation, which states that "you should expect a stepwise, always-increasing behavior for statistical significance." (Optimizely 2025). Optimizely customer support explained that this is due to the time bucketing algorithm

[12] See Data Collection Statement for details on Optimizely

and "stat reset," when the Stats Engine spots a seasonality and drift in the conversion rates.

2. The data for Obs and Obs-BYGG was collected by Kameleoon, an A/B testing platform.[13] The automated analysis provided by Kameleoon's frequentist algorithm matches the lift and p-values computed in Excel and shown in Table 1.

3. There is no indication of SRM (Sample Ratio Mismatch) in any of the experiments (Fabijan et al. 2019; Kohavi, Tang, and Xu 2020).

4. Looking at daily graphs of users and conversion rates, we do not see any unexpected anomalies. For Obs and Obs-BYGG there was a spike in conversion rates on the November 11, 2024, holiday, and smaller spikes on weekends, which Coop said is expected.

We report summary statistics for the experiments without providing anonymized raw data for several reasons:

1. Historical precedent shows that "anonymized" raw data can create substantial legal and reputational exposure. Here are four well-known examples that illustrate the risk:

    a. AOL released query logs that anonymized users, yet the New York Times was able to locate individuals (Wikipedia contributors, AOL search log release 2025). AOL's CTO resigned and two employees who were involved were fired (Zeller 2006). A lawsuit accused AOL of violating the Electronic Communications Privacy Act, which AOL settled for $5 million (Davis 2013). Business 2.0 Magazine on CNNMoney ranked the release of the data as #57 of its "101 Dumbest Moments in Business" for 2007 (CNN 2018).

    b. Netflix released anonymized movie ratings for the Netflix Prize. They were subsequently sued for privacy invasion (Singel 2009). Law professor Paul Ohm called it a "multi-million-dollar privacy blunder" (Ohm 2009). Netflix settled and canceled the follow-on "Netflix Prize 2" (Ohm 2010).

    c. The NYC Taxi and Limousine Commission released a dataset of 173 million taxi trips with anonymized medallion numbers but precise timestamps of pickup/drop-off locations. Anthony

[13] See Data Collection Statement for details on Kameleoon

Tockar was able to de-anonymize the data by cross-referencing trip times with paparazzi photos, showing celebrity trips and tipping amounts (Tockar 2014).

d. Fitness tracking app Strava released a heat map of aggregated user activity, but it revealed the locations and layouts of secret U.S. military bases as mapped out by the tracked jogging routes of soldiers (Hern 2018).

2. Data with timestamps can often be joined with external sources and fingerprinting techniques can expose identities. For example, a user might tweet "Just bought my SeaWorld tickets," and that information can be joined with the raw data to reveal the user's identity.

3. Timestamps reveal commercially sensitive operational patterns. Websites' hourly and daily traffic, weekend vs. weekday traffic, and holiday surges are competitively sensitive and can be tied to campaigns and confidential operational constraints.

4. Time series data reveals retention, repeat-visit, and attrition behavior. User journeys across sessions can be reconstructed and leak sensitive business information on funnels and customer patterns.

5. Additional attributes about users, such as location and demographics, increase fingerprinting risk. Location, device, browser, and landing pages can allow others to identify users.

Given these concerns, the *Trustworthy A/B Patterns* project's legal agreement was designed to support publication of approved summary statistics by participating companies. The three expert volunteers in the project (Kohavi, Linowski, and Vermeer) operated under an NDA (Non-Disclosure Agreement) and were able to inspect additional time series information and, in some cases, raw data to validate correctness. Additionally, for each participating company, the key person involved in the project is also a co-author of this paper. See Appendix C for additional details.

Our replications are not reproductions as defined in *Reproducibility and Replicability in Science* (National Academies of Sciences, Engineering, and Medicine 2019) because we are using different samples and different websites. In the terms suggested by Urminsky and Dietvorst (2024), these are theory-test replications: we implemented similar interventions and used the same estimation method, but on different websites. A direct or exact replication is impossible for Study 2, as the startup website used in BAC no longer exists. More important, external generalizability of a theory can be more useful, as highlighted by Lynch et al. (2015).

There are some differences between our replications and the A/B test (Study 2) in BAC, including direction (square to round or round to square), geography (US vs. Norway) and what click-through rate is applied to. These are summarized in Table 2 below. In addition, the last two rows show the experiment power to detect a positive treatment effect of 2%, which was recommended in the original *Trustworthy A/B Patterns* project (Kohavi, Linowski, and Vermeer 2024), and 0.21%, which is our best estimate of the true effect given the meta-analysis above. Note that even with our three large sites, we are underpowered to detect the tiny effect that is our best estimate of the true delta. Sites that do not have many millions of users in an experiment period will therefore not be able to run a properly powered A/B test to evaluate whether rounded vs. square corners are better for them.

**TABLE 2: Different factors in the experiments**

| Experiment factor | BAC | SeaWorld | Obs | Obs-BYGG |
|---|---|---|---|---|
| Treatment round or square (forward or reverse experiment) | T: round | T: round | T: square | T: square |
| Geography | US | US | Norway | Norway |
| Primary metric | CTR on the "Contact Us" button | CTR on any button | CTR on add-to-cart | CTR on add-to-cart |
| Power for positive 2% MDE | 3.0% | 100.0% (rounded) | 99.7% | 94.9% |
| Power for positive 0.21% MDE | 2.6% | 39.2% | 7.1% | 5.7% |

*Note*: The power shown is for detecting a positive treatment effect (right-tail; 2.5% is the type-I error rate)

Our replications resulted in very small effects: two orders of magnitude smaller than the initial result. We do not believe the differences in the table are material, nor do we believe that some mediator could explain the dramatically different lift estimates. We believe it is more likely that the previously published results are an occurrence of the winner's curse, which manifests as significant exaggeration of the treatment effect in under-powered studies (Button et al. 2013; Gelman and Carlin 2014). We now examine whether that initial result used too small a telescope.

# Small telescope—are the original studies too small?

Are our results inconsistent with BAC's reported results? Statistically, they *are* inconsistent (see Appendix A), but this comparison is not the right question to ask. In Simonsohn (2015), the author points out that results that are "barely significant (e.g., p> 0.025 rather than p < 0.025) …have confidence intervals that nearly touch zero." For example, BAC's Study 2 has a p-value of 0.037 with a very wide 95% treatment effect confidence interval of [0.27%,7.71%] in absolute percentage points,[14] and a massively large relative effect confidence interval of [3.13%,146.8%] around the point estimate of 55%. Note that computing confidence intervals on relative effects (lift) requires use of Fieller's theorem, as we are dividing two random variables (Fieller 1954; Willan and Briggs 2006; Kohavi et al. 2009).

With the small telescope approach, we ask not whether the results of the replication are consistent with the *results* of the original study, but whether they are consistent with its *design*. In other words, was the original study sufficiently powered to be informative? Simonsohn (2015) uses the following analogy:

> Imagine an astronomer claiming to have found a new planet with a telescope. Another astronomer tries to replicate the discovery using a larger telescope and finds nothing. Although this does not prove that the planet does not exist, it does nevertheless contradict the original findings, because planets that are observable with the smaller telescope should also be observable with the larger one. It is generally very difficult to prove that something does not exist; it is considerably easier to show that a tool is inadequate for studying that something. With a small-telescopes approach, instead of arriving at the conclusion that a theoretically interesting effect does not seem to exist, we arrive at the conclusion that the original evidence suggesting a theoretically interesting effect exists does not seem to be adequate [that is, the original telescope was too small]. (560)

To determine whether the sample size in the original study was too small, we ask what true effect size would have given the original study 33% power, call that $d_{33}$. If the replication's effect size is significantly smaller than $d_{33}$, we reject the hypothesis that the studied effect was large enough to have been detectable with the original sample size.

Figure 7 shows the relative effect sizes of the original field study (Study 2), the two lab studies (Study 3 and Study D) and our three replications with their 95% confidence intervals. Given that the 95% confidence intervals are tiny and far from

[14] Appendix A provides additional details based on feedback from a reviewer, who pointed out that for small counts, it is appropriate to apply Yates continuity correction, which results in a p-value of 0.049 and a confidence interval that is even closer to zero.

$d_{33}$, the effect with 33% power line, we conclude that the original study was too small.[15]

Simonsohn (2015) suggests that replication sample sizes be 2.5 times the original sample. Our replications are over *two thousand times* larger than the original study.

**Figure 7:** Relative effect sizes for the three original studies by BAC and our three replications with 95% confidence intervals.

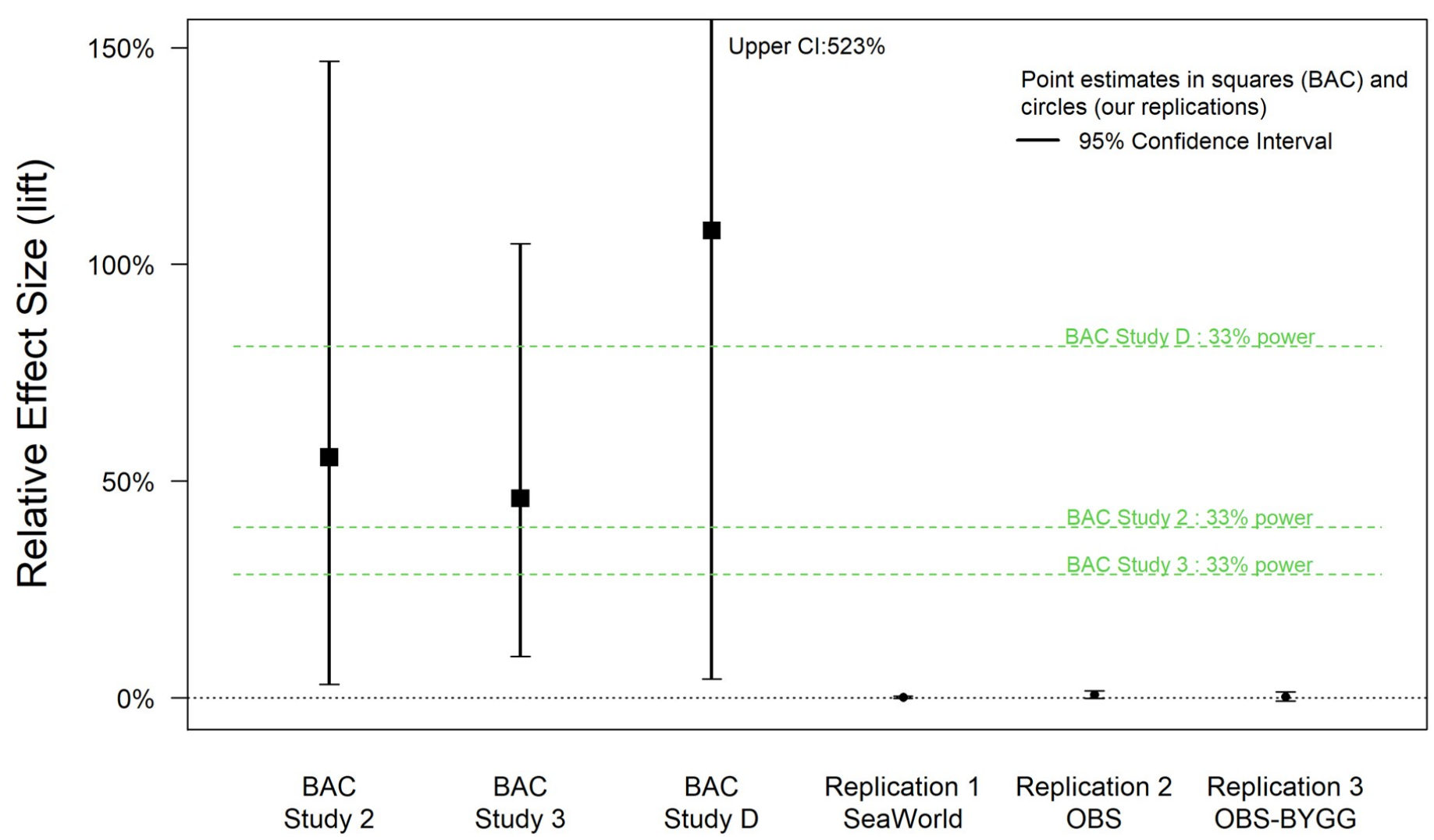

*Note*: Because the 33% power lines do not intersect the confidence intervals for the three replication studies, the original studies are considered too small.

## Additional industry replications

Evidoo (Jongejan 2025) claims to be the world's largest online A/B test database for e-commerce, based on over 20 years of conducted A/B tests. In December 2024, they ran two experiments on two different sites, changing the add-to-cart button from square to rounded corners on mobile devices. Conceptual designs are shown in Figure 8.

[15] R code producing the chart is available at https://bit.ly/roundedCornersComputations under the small telescopes tab.

**Figure 8:** Changing the Add-to-Cart button from square to round

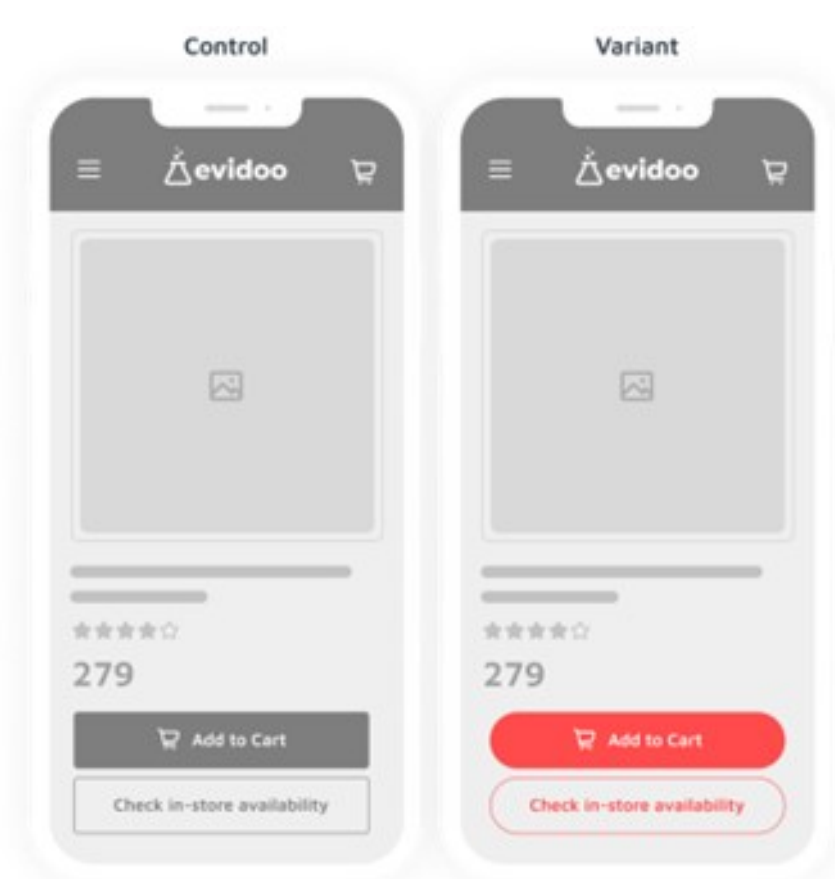

Due to legal consulting agreements, they are unable to share the company names, but the results are summarized in Table 3 below, based on https://www.evidoo.io/best-practices/1380/ and email exchanges with the CEO. We replicated their computations in https://bit.ly/roundedCornersComputations under the Evidoo tab.

**TABLE 3: Two A/B Replication Tests by Evidoo**

| | N control | N treatment | Conversion Control | Conversion Treatment | Lift | p-value |
|---|---|---|---|---|---|---|
| Client 1 | 84,120 | 84,336 | 12.3% | 12.4% | 0.68% | 0.60 |
| Client 2 | 83,126 | 83,041 | 12.2% | 12.2% | 0.07% | 0.96 |

BAC's Study 2 was based on 919 visits and reported an effect size of 55%. With over 160,000 users in each experiment, the sample sizes in these replications are over 170 times as large. They have about 97% power to detect a 5% change, but only about 33% power to detect a 2% change, which was recommended for this pattern in the *Trustworthy A/B Patterns* project (Kohavi, Linowski, and Vermeer 2024). The p-values of 0.60 and 0.96 are far from 0.05, and the effect sizes are two orders of magnitude smaller than the 55% originally reported.

## Discussion

Carl Sagan famously wrote, "Extraordinary claims require extraordinary evidence." We suspected that BAC's reported effects were implausible and conducted three highly powered replication experiments to investigate. Our results indicate that, in all our experiments, the treatment effect is not statistically significantly different from zero. If there is a true effect, it is about two orders of magnitude smaller than the previously reported estimate.

While there are some differences in the design of our experiments from the original study, these are theory-test replications (Urminsky and Dietvorst 2024) that evaluate the claim that "Rounded Shapes Enhance Online Click-Through Rates" in the title of the BAC paper. We think the time-reversal heuristic (Gelman 2016) is appropriate to apply here. Had our three replications been published first, showing that the impact of rounded corners is consistent with the null hypothesis of no effect, would readers be inclined to believe a study with a sample size 2,000 times smaller?

The p-values reported by BAC's three studies to evaluate click-through rate are: 0.037, 0.011, and 0.045 for Study 2, Study 3, and Study D. If an experiment has 80% power and the true effect equals the MDE, then the expected p-value is 0.005, corresponding to Z of 2.8 in a two-tailed test (van Belle 2008). Bartoš and Schimmack (2022) highlight that borderline results around the alpha cutoff of 0.05 for p-value are a strong indication that the results will not replicate well. Kohavi et al. (2022) show that across multiple organizations running A/B tests at scale, the median success rate of experiments is 10%, so using an alpha threshold of 0.05 implies that a statistically significant result will be a false positive 22% of the time, even if properly powered at 80%. Benjamin et al. (2017) proposed reducing the alpha threshold to 0.005 and calling results with higher p-values "suggestive," requiring the accumulation of additional evidence.

In "Estimating the Reproducibility of Psychological Science" (Open Science Collaboration 2015), replications of 100 experimental and correlational studies were conducted using high-powered designs. The mean effect size of the replication effects was half the original, and only 36% of replications had significant results ($p < 0.05$), compared to 97% of the original studies.

Button et al. (2013, 372) emphasized how low statistical power reduces the probability that a statistically significant result reflects a true effect. They claim that "low power therefore has an ethical dimension—unreliable research is inefficient and wasteful." Gelman and Carlin (2014) show the expected exaggeration in low powered experiments. Using an MDE of 2%, which is what was recommended in the *Trustworthy A/B Patterns* project (Kohavi, Linowski, and Vermeer 2024) given similar experiments, the power of BAC's A/B test (Study 2) to detect a positive change is just 3% (when the type-I error for the right tail is 2.5%). At such low power, the degree of exaggeration (Gelman and Carlin 2014), assuming no sign-error, is expected to be about 28 times! Using our meta-analysis best estimate of 0.21%, the power of BAC's A/B test to detect a positive change is just 2.55%, implying an expected exaggeration of over 200.[16]

Szucs and Ioannidis (2017) argue that use of Null Hypothesis Significance Testing (NHST) should require power calculations. There were no power calculations in BAC's paper to justify the sample size.

## The cost of exaggerated claims

In NHST, a p-value is computed under the assumption that the null hypothesis is true. Given a prior on the null hypothesis, Bayes Rule can then be used to compute the False Positive Risk, or the probability that the statistically significant result is a false positive (Colquhoun 2017; Wacholder et al. 2004; Ioannidis 2005; Kohavi et al. 2014; Benjamin et al. 2017).

Our experience, and the set of similar experiments shown above that had estimated treatment effects much smaller than 55%, led us to be very skeptical of the BAC's results, given the borderline p-value of 0.037. The large replications indeed showed that, even if there is an effect, it is estimated to be two orders of magnitude smaller than 55%. The theory that rounded shapes materially increase click-through rate is very fragile.

If an experiment the size of our SeaWorld replication had a true 55% lift and an observed value of 55%, the p-value would be less than 2.2e-16,[17] a tiny number that would convince us of the treatment effect even if our prior probability was 1 in a million (1e-6).

We share two examples of exaggerated claims and their lasting negative impact:

1. Power posing. In 2010, a tiny study (Carney, Cuddy, and Yap 2010) was executed with n=42 students from Columbia University, who were randomly split into Control and Treatment. The claim was that "High-power posers [in the Treatment group] experienced elevations in testosterone, decreases in cortisol, and increased feelings of power and tolerance for risk…by assuming simple 1-min poses, a person can embody power and instantly become more powerful, which has real-world actionable implications." This was followed up by a TED Talk by the second author (Cuddy 2012), who provided a "free, no-tech life hack…change your posture for two minutes." She explained that they found "on testosterone…. high-power people experience about a 20-percent increase, and low-power people experience about a 10-percent decrease…on cortisol... High-power people experience about a 25-percent decrease, and the low-power people experience about a 15-percent increase." The audience was asked "to share the science" (Ibid.).

[16] The power computations are available at https://bit.ly/roundedCornersComputations under the "Power from N" tab.

[17] See https://bit.ly/roundedCornersComputations under the Proportions test tab

In 2015, Cuddy published a 300-page book (Cuddy 2015) that became a New York Times bestseller. That year, a relatively large replication of Carney et al.'s study failed (Ranehill et al. 2015). This was followed by analysis by Data Colada (Simmons and Simonsohn 2015; Simmons and Simonsohn 2017), which concluded that "evidence for the basic effect seems too fragile to search for moderators or to advocate for people to engage in Power Posing to better their lives."

In 2016, the lead author wrote "I do not believe that 'Power Pose' effects are real …The evidence against the existence of power poses is undeniable" (Carney 2016). Yet the TED talk has been played over 74 million times. Using the internet archive, we estimate that there were 2.1 million plays in the year starting October 17, 2024, or over 5,700 plays per day (Kohavi 2025).

Every day, thousands of people continue to hear the amazing story of power posing, which is based on an incorrect claim from a tiny study. Over half a million people bought Cuddy's 2015 book.

2. Priming. In the original priming study (Bargh, Chen, and Burrows 1996), n=30 undergraduates from New York University (a tiny sample) were split randomly into Control and Treatment. Those in the Treatment were primed with elderly-related words, and they walked more slowly out of the lab than those in the Control group. The article has been cited over 6,800 times according to Google Scholar, and has been cited in many books, including *Thinking Fast and Slow* by Nobel prize winner Daniel Kahneman (2011, 40), who wrote: "Disbelief is not an option. The results are not made up, nor are they statistical flukes. You have no choice but to accept that the major conclusions of these studies are true." When confronted with evidence that the replicability index of the chapter is under 20% (Schimmack 2020), Kahneman responded with "I placed too much faith in underpowered studies…authors who review a field should be wary of using memorable results of underpowered studies as evidence for their claims" (Kahneman 2017). The damage was that "millions of people had been informed by a Nobel Laureate that they had 'no choice' but to believe in those priming studies" (Ritchie 2020).

BAC's paper was cited by 23 articles around a year after publication according to Google Scholar. They were featured in the *Tampa Bay Times* (Berdychowski 2023) and appeared in multiple postings on LinkedIn. Brandolini's law says (in its polite version) that the energy to debunk far exceeds the energy to make a claim (Williamson 2016). We hope this article will not only correct this specific exaggeration but also help people understand the criticality of establishing a reasonable MDE and doing power calculations.

## Conclusion

Several years ago, a magazine cover story, "How Science Goes Wrong" (The Economist 2013) reaffirmed the simple idea that underpins science: "Trust but verify." The article pointed out that "Negative results now account for only 14% of published papers…Yet knowing what is false is as important to science as knowing what is true . . . replication is hard and thankless. Journals, thirsty for novelty, show little interest in it" (Ibid.).

Our results showed estimated lifts of just 0.16%, 0.29%, and 0.73%, and two industry replications in Evidoo (Jongejan 2025) had estimated lifts of 0.68% and 0.07%. None of the experiments allowed rejecting the null hypothesis of no effect at the industry standard p-value threshold (alpha) of 0.05, and these experiments were run on relatively large websites. We believe that BAC's reported lift of 55% for their A/B test was overstated by two orders of magnitude, a typical consequence of low-powered published results.

These findings align with broader concerns about the reliability of low-powered studies and their tendency to produce exaggerated or extreme results (Button et al. 2013; Gelman and Carlin 2014). The absence of power calculations in the original study further highlights the importance of rigorous experimental design and careful interpretation of results.

Scientific progress depends on scrutiny, replication, and robust methodologies. By applying high-powered experiments and meta-analysis, we provide a more accurate estimate of the effect—if there is any—of rounded shapes on click-through rates. Our results highlight the importance of statistical power and caution against accepting exaggerated findings without sufficient and trustworthy evidence.

# Appendix A

BAC reported a p-value of 0.037 for Study 2, but using the two-sample test for equality of proportions in R (R Core Team 2025) gives a higher p-value of 0.049. Specifically, running `prop.test(x=c(53,32), n=c(474,445))` returns a p-value of 0.049, which is higher than the p-value of 0.037. We managed to replicate the p-value and $\chi^2$ reported by BAC by modifying R's default by disabling Yates continuity correction, that is, by running `prop.test(x=c(53,32), n=c(474,445), correct=FALSE)`.

The R code and results for all experiments using the proportion test with and without Yates correction are available at https://bit.ly/roundedCornersComputations under the "Proportion test" tab.

As discussed in the Small Telescope section, looking at confidence intervals to determine whether a replication contradicts the original experiment is inappropriate, as the confidence interval for a borderline p-value is always going to be close to zero, implying that almost any positive result would be consistent with the confidence interval.

Nonetheless, here are the confidence intervals for Study 2 (without Yates continuity, matching the BAC results):

1. The 95% confidence interval for the absolute effect in Study 2 is [0.27%, 7.71%]. Our absolute treatment effects are 0.07%, 0.07%, and 0.02%, so all of them are outside the 95% confidence interval.

2. The 95% relative confidence interval, using Fieller's theorem, is [3.13%, 146.8%]. Our relative treatment effects (lifts) are 0.16%, 0.73%, and 0.29%, so all of them are outside the 95% confidence interval.

# Appendix B

The Obs and Obs-BYGG experiments had a second treatment implemented, which had slightly rounded corners, as shown below in Figure 9.

**Figure 9:** A second treatment for each of the Coop experiments with slightly rounded corners

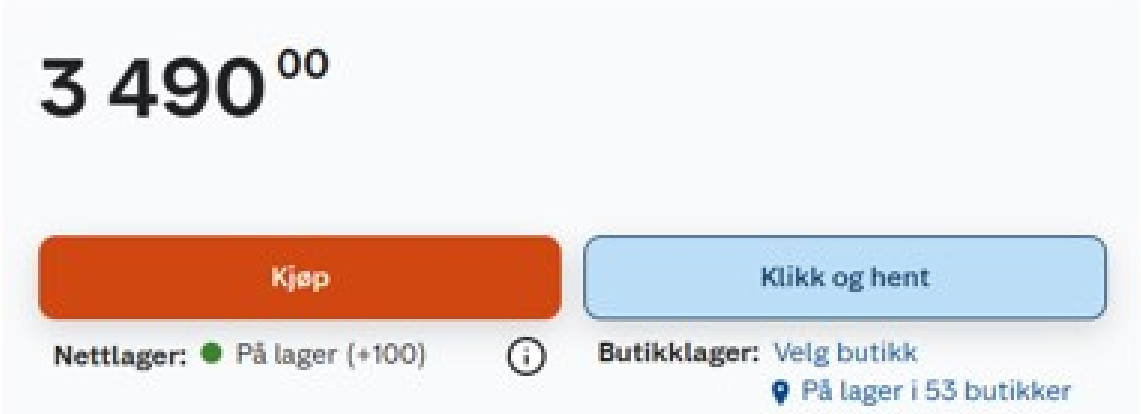

Table 4 shows the summary of results, including the two additional treatments with slightly rounded corners, denoted by "(SR)."

**TABLE 4: Summary of results including the slightly rounded treatments, denoted by SR.**

| Study | N control | N treatment | Conversion Control | Conversion Treatment | Lift | p-value |
|---|---|---|---|---|---|---|
| SeaWorld® Orlando | 1,448,041 | 1,448,066 | 47.13% | 47.21% | 0.16% | 0.20 |
| Obs-BYGG | 1,126,132 | 1,124,100 | 5.43% | 5.45% | 0.29% | 0.60 |
| **Obs-BYGG (SR)** | 1,126,132 | 1,124,002 | 5.43% | 5.46% | 0.54% | 0.33 |
| Obs | 977,499 | 976,653 | 10.07% | 10.14% | 0.73% | 0.09 |
| **Obs (SR)** | 977,499 | 976,966 | 10.07% | 10.14% | 0.72% | 0.09 |

*Note*: The Control is shared for both treatments, hence the exact same Control count and Control conversion

We are sharing this for completeness, but these additional treatments do not represent completely independent replications (e.g., they share the same control group). Also, as Vosgerau et al. (2019) noted, internal meta-analyses are often invalid. Finally, their UX design does not match the treatment in the original paper.

# Appendix C

This appendix details the data collection and analysis for the reported experiments.

## SeaWorld Replication Experiment

One of the co-authors (Ravikiran Rajagopal) supervised the data collection for the A/B tests conducted using the Optimizely Web Experimentation platform from May–June 2025. The same author continuously oversaw all data collection activities within the Optimizely tool throughout the experiment period.

Data was collected via the Optimizely Web Experimentation JavaScript snippet, which is implemented on the client side of the website. This snippet enables the collection of event and user data necessary for experimentation and analysis. See: https://support.optimizely.com/hc/en-us/articles/4410284311565-Optimizely-Web-Experimentation-JavaScript-snippet.

Details of the specific data collected by Optimizely can be found in the official data specification documentation. See: https://docs.developers.optimizely.com/experimentation-data/docs/experimentation-events-data-specification.

### Data Security and Processing

Collected data is processed and transferred in accordance with high-security standards. Optimizely encrypts customer data both in transit and at rest, ensuring data privacy and security throughout its lifecycle. Further details on encryption practices are available in Optimizely's security documentation. See: https://support.optimizely.com/hc/en-us/articles/4410289778701-How-Optimizely-Experimentation-encrypts-customer-data-at-rest.

### Data Retention

Optimizely maintains a clear data retention policy, outlining how long data is stored and when it is deleted, as described in their data retention policy and FAQs. See: https://support.optimizely.com/hc/en-us/articles/8210191986061-Data-retention-policy-and-FAQs.

### Data Validation

Two of the authors (Ron Kohavi and Ravikiran Rajagopal) conducted several validation checks for sample ratio mismatch, user contamination, filter condition checks, outlier and distribution analysis, and instrumentation effect to ensure the integrity and reliability of the experimental results.

## Coop Replication Experiments

One of the co-authors (Joachim Furuseth) supervised the data collection for the two A/B tests conducted using an experimentation tool called Kameleoon from September to December 2024.

The same author continuously supervised the data collection within the Kameleoon tool throughout the entirety of the experiment period.

The data was collected through a snippet of code (.js file) implemented on the client-side of the website.

Details of the data collected can be found on Kameleoon technical documentation (https://developers.kameleoon.com/privacy-and-compliance/data-collection-back-end/#list-of-collected-data).

Data is then processed and transferred with high-security standards (encryption both in transit and at rest; details on Kameleoon security portal: https://security.kameleoon.com/?itemName=data_security) to servers located within the EU in France, Germany, and Finland.

Another author (Ron Kohavi) ran several validation checks for sample ratio mismatch and user contamination.

## About the Authors

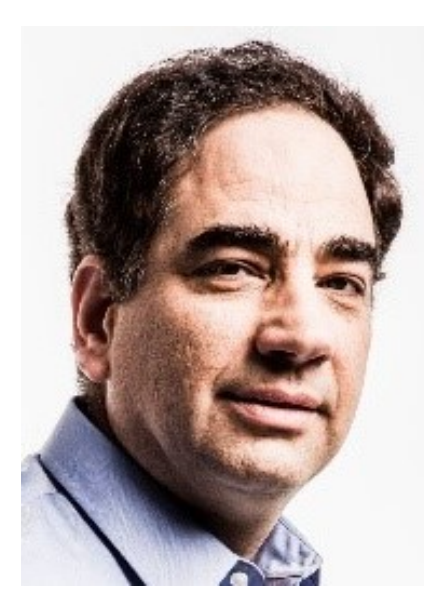

**Ron Kohavi** received his Ph.D. in Machine Learning from Stanford University and his BA from the Technion, Israel. He now consults and teaches A/B testing courses on Maven. He was Vice President and Technical Fellow at Airbnb, a Microsoft Technical Fellow and Corporate VP at Microsoft, and Director of Data Mining and Personalization at Amazon.com. He received the Individual Lifetime Achievement Award for Experimentation Culture in 2020. He co-authored the book *Trustworthy Online Controlled Experiments: A Practical Guide to A/B Testing* with Diane Tang and Ya Xu. His papers have over 65,000 citations. His email address is ronkohavi@outlook.com.

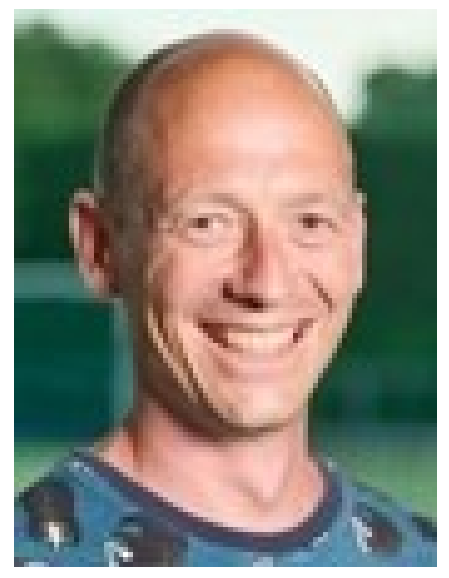

**Lukas Vermeer** holds an MSc in Computing Science from Utrecht University with a minor in Artificial Intelligence. He is currently Senior Director of Product at Vistaprint and occasional keynote speaker on the topic of online experimentation. Previously, he was Director of Experimentation at Booking.com. He has received two Experimentation Culture Awards and his impact was recognized in the Harvard Business Review story "Building a Culture of Experimentation" (March–April 2020 issue). Lukas has co-authored academic papers on online experimentation and spoken at over 30 conferences, including KDD and SIGIR. His email address is lukas@lukasvermeer.nl.

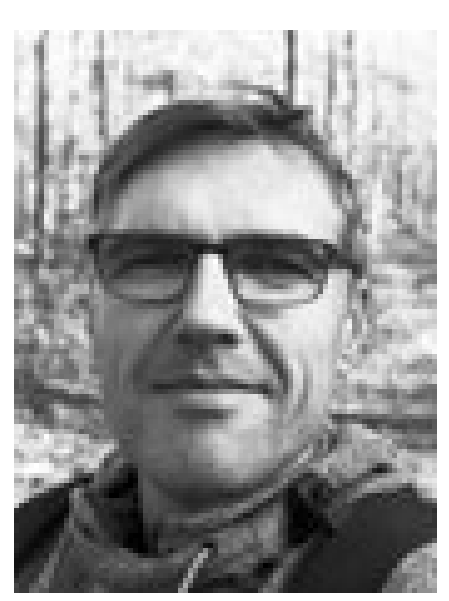

**Jakub Linowski** practices UI design (since ~2000) and experiment design (since 2014). He founded GoodUI.org, an active repository of experiments aimed at uncovering generalizable UI patterns and encouraging replication. Over the years, Jakub has been hired by, consulted for, collaborated with, shared results with, coached, and learned from 100+ experimentation teams, including: Microsoft, Thomasnet, Booking, Backstage, Yummly, Elevate, Designlab, Reverb, VivaReal, ZAP, Volders, Rollbar, Kenhub, Expert Institute, Drip Agency, FINN, 3D Hubs, and Metro. Jakub holds two design degrees, including an MSc in Design for Interaction from Delft University of Technology (Netherlands). His email address is jlinowski@gmail.com.

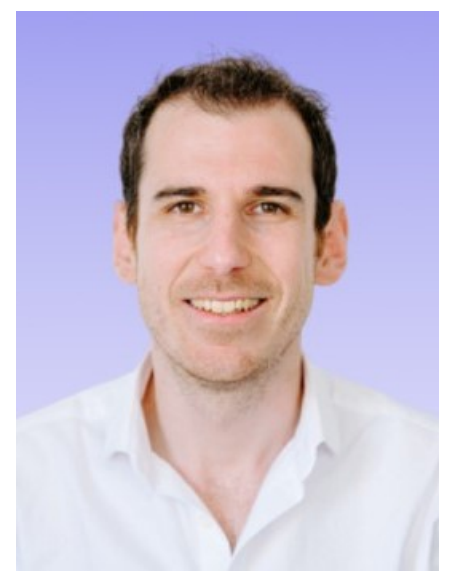

**Fabrice Boisseranc** is a Senior Solution Engineer at Kameleoon, the experimentation platform that is used by Coop. After an engineering degree in Computer Science and 10 years of experience in the SaaS Industry, he has been working in the experimentation field at Kameleoon for 4 years and is now an expert of the Kameleoon platform, with a strong focus on analytics and statistics. In his role, he acts as a pivot between the product team and the other teams within the organization (Sales, Customer Success, Marketing and Partnership). His email address is fboisseranc@kameleoon.com.

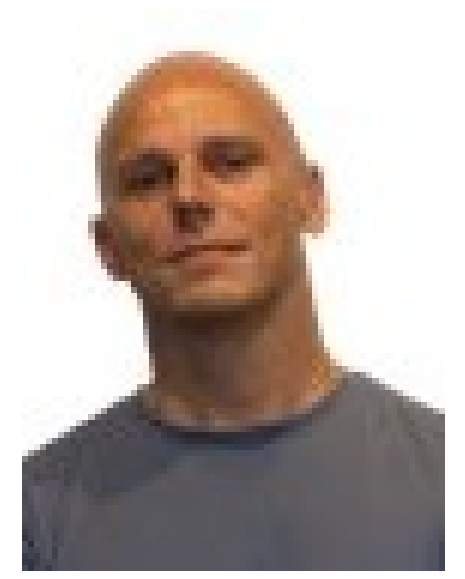

**Joachim Furuseth** is an experimentation and CRO leader with more than 11 years of hands-on experience building and scaling data-driven optimization programs. He has worked extensively from the agency side with several of Norway's largest brands, helping organizations mature their experimentation practices, analytics foundations, and decision-making culture. Joachim was the founder of Right Brain, Norway's first and only agency fully dedicated to CRO and experimentation at the time. He currently works at Coop, one of Norway's top 15 largest companies with more than 26,000 employees, where he leads experimentation and insights within eCommerce. His expertise spans large-scale A/B testing, analytics, experimentation strategy, and embedding experimentation as a core business capability. His email address is joachim.furuseth@coop.no.

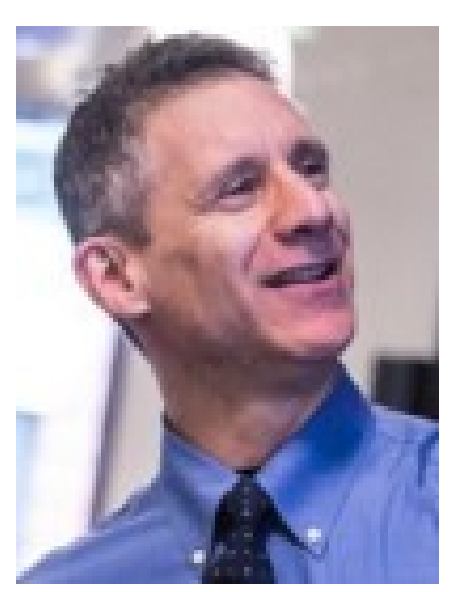

**Andrew Gelman** is a professor of statistics and political science at Columbia University. He has received the Outstanding Statistical Application award three times from the American Statistical Association, the award for best article published in the American Political Science Review, the Mitchell and DeGroot prizes from the International Society of Bayesian Analysis, and the Council of Presidents of Statistical Societies award. His books include *Bayesian Data Analysis* (with John Carlin, Hal Stern, David Dunson, Aki Vehtari, and Donald Rubin), *Teaching Statistics: A Bag of Tricks* (with Deborah Nolan), *Data Analysis Using Regression and Multilevel/Hierarchical Models* (with Jennifer Hill), *Red State, Blue State, Rich State, Poor State: Why Americans Vote the Way They Do* (with David Park, Boris Shor, and Jeronimo Cortina), *A Quantitative Tour of the Social Sciences* (co-edited with Jeronimo Cortina), *Regression and Other Stories* (with Jennifer Hill and Aki Vehtari), and *Active Statistics* (with Aki Vehtari and Richard McElreath). His email address is ag389@columbia.edu.

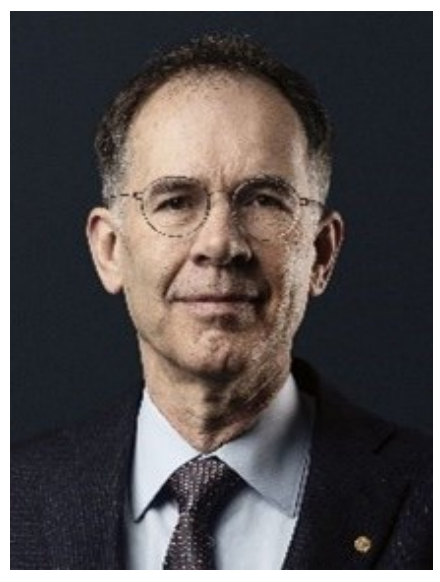

**Guido Imbens** received his Ph.D. from Brown University in 1991. He is currently Professor in the Graduate School of Business and the Economics Department and director of Stanford Data Science at Stanford University. He previously held tenured positions at UCLA, UC Berkeley and Harvard University. He has published on causal inference in leading economics and statistics journals. He served for six years as the main editor of *Econometrica*. He is a coauthor with Donald Rubin of *Causal Inference for Statistics, Social and Biomedical Sciences*. He is a fellow of the American Academy of Arts and Sciences and an elected member of the National Academy of Sciences. He is a co-recipient of the 2021 Sveriges Riksbank Prize in Memory of Alfred Nobel. His email address is imbens@stanford.edu.

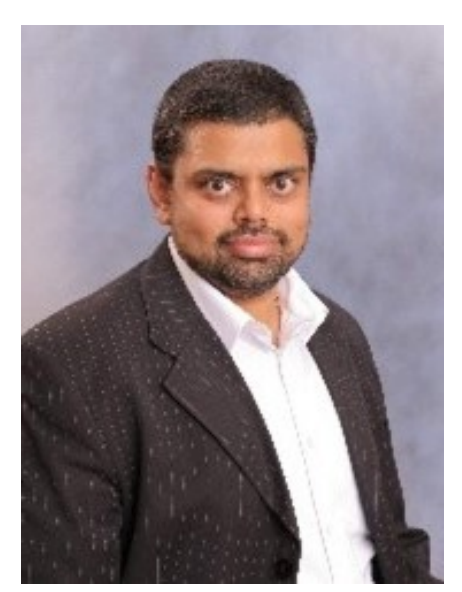

**Ravikiran Rajagopal** is a product and experimentation leader focused on improving customer outcomes through rigorous A/B testing and data-driven decision-making. His work centers on identifying friction in high-traffic digital journeys and validating solutions through controlled experimentation at scale. At SeaWorld Parks & Entertainment, he built and scaled an enterprise experimentation program across multiple consumer platforms, embedding testing into core product and growth decisions. This work was recognized externally with an award for best experimentation program. Earlier in his career at Prudential Financial, he focused on improving customer experience and adoption in regulated financial products through product design and platform modernization. His email address is ravikiran.rajagopal@gmail.com.